# Nonlinear Boundaries in Quantum Mechanics

Arthur Davidson
ECE Department
Carnegie Mellon University,
Pittsburgh, PA 15213
artdav@ece.cmu.edu

**Abstract:** Based on empirical evidence,[1] quantum systems appear to be strictly linear and gauge invariant. This work uses concise mathematics to show that quantum eigenvalue equations on a one dimensional ring can either be gauge invariant or have a linear boundary condition, but not both. Further analysis shows that non-linear boundaries for the ring restore gauge invariance but lead unexpectedly to eigenfunctions with a continuous eigenvalue spectrum, a discreet subset of which forms a Hilbert space with energy bands. This Hilbert space maintains the principle of superposition of eigenfunctions despite the nonlinearity. The momentum operator remains Hermitian. If physical reality requires gauge invariance, it would appear that quantum mechanics should incorporate these nonlinear boundary conditions.

## Introduction

To start, the theory will be reviewed showing that momentum in an infinite 1 dimensional system (not a ring) with boundaries at infinity is fully linear and gauge invariant. The momentum eigenvalue equation will be written explicitly in gauge dependent form, and it will be shown that identical eigenvalues result, independent of gauge. The linearity of this infinite one dimensional system with effectively no boundary conditions will also be apparent. These results, of course, are expected from the linearity and gauge invariance of standard quantum mechanics.

Next, a finite 1 dimensional ring will be considered, again with a momentum eigenvalue equation that can be written in gauge dependent form. But now the wave function is finite all the way around the ring, and the way it interacts with itself at the boundary must be described. It will be shown that linear boundary conditions lead to different eigenvalues for different gauges.

It will then be shown that giving up linearity at the boundary restores gauge invariance. The gauge invariant eigenvalues will form a continuous spectrum of real numbers. However, a subset of the eigenfunctions will be allowed by the nonlinear boundary conditions to form a Hilbert space and be superposable. This subset turns out to be within a phase factor of the complete orthonormal set allowed by linear boundaries, if the issue of gauge invariance is ignored. Thus the advantage of nonlinear boundary conditions for the one dimensional ring is that essentials of Hilbert space and the superposition principle are maintained along with gauge invariance and Hermiticity of the momentum operator.

The eigenfunctions associated with the continuous spectrum are generally not orthonormal or superposable. Thus it cannot be simply assumed that all eigenstates of a system are superposable: it may be that Schrodinger's cat is never found in superposition because the two states of the cat are not orthonormal and not superposable. Likewise, superposability cannot be assumed for all eigenstates or qubits of a quantum computer.

## The Infinite One Dimensional Case

Consider one coordinate dimension extending toward plus and minus infinity. The gauge dependent momentum operator can be written:

$$p_{op} = -i\frac{\partial}{\partial x} - k \ . \qquad \text{Eq. 1}$$

Units have been chosen so that $\hbar = 1$, $k$ is an arbitrary real constant different for each gauge and $x$ represents the coordinate.

The gauge dependent eigenfunction can be written

$$\psi(x) = Ae^{i(n+k)x} \qquad \text{Eq. 2}$$

where A is a constant determined by normalization. As is well known in standard quantum mechanics, an operator applied to one of its eigenfunctions should yield a real constant eigenvalue multiplying the same eigenfunction: Thus

$$\left(-i\frac{\partial}{\partial x} - k\right)\left(Ae^{i(n+k)x}\right) = n\left(Ae^{i(n+k)x}\right) \qquad \text{Eq. 3}$$

so that the real number n is the gauge independent eigenvalue. The number k, which is gauge dependent, cancels out of the eigenvalue. Note that boundary conditions play no role in this one dimensional problem extending far from the origin. As usual in standard quantum mechanics, the eigenfunctions extend over the whole domain, and n forms a continuous spectrum of real values. Since Eq. 3 is linear and homogeneous in the eigenfunction, and there are no effective boundary conditions other than normalization, this system, including the operator and eigenfunction together, is linear. All eigenfunctions are part of the Hilbert space.

This system has a gauge dependent momentum operator, and gauge dependent eigenfunctions. However, when the operator and an eigenfunction are put together the result is a gauge independent eigenvalue. Since measurement theory holds that the eigenvalue should correspond to possible measured momentum values, the overall quantum system here is properly gauge invariant.

Therefore, it has been confirmed that this infinite one dimensional system is linear and gauge invariant. All the eigenfunctions form the usual orthonormal Hilbert space, and all can be superposed with one another in any linear combination.

## The Finite One Dimensional Ring with Linear Boundary Conditions

Equations 1, 2, and 3 should apply to the finite one dimensional ring as well, except that the eigenfunctions should now also satisfy a definite boundary condition. Standard quantum mechanics and most physicists' starting assumption would be that these boundary conditions are linear. That is, even though the coordinate may have a discontinuity somewhere due to the nature of the ring geometry, the wave function should join onto itself smoothly everywhere, including at the boundary indicated by the coordinate discontinuity. Thus, if the coordinate is chosen such that x extends from $-\pi$ around the ring to $\pi$ then the eigenfunction in Eq. 2 must have $(n + k) = m$, where m is an integer. This is needed for the wave function to be periodic and to join smoothly and linearly with itself. The eigenvalue spectrum is discreet but gauge dependent. That is, the eigenvalue n corresponding to equation 3 will be $n = m - k$. This eigenvalue depends explicitly on k, and so is dependent on the gauge.

There are only two differences between the infinite line and the ring: the ring has a finite coordinate, and has a definite boundary condition. The finite ring domain is the system under investigation, so the salient thing to change to restore gauge invariance is the boundary condition.

## The Finite One Dimensional Ring with Nonlinear Boundary Conditions

What boundary condition would restore gauge invariance of the momentum eigenvalue? The obvious thing to try is to demand smoothness at the coordinate discontinuity not of Ψ the wave function itself, but of the probability density ψ*ψ and the probability current density ψ*ψ(∇Φ), where ∇Φ is the gradient of the gauge dependent wave function phase. Aside from being distinctly nonlinear, these functions of the wave function would allow the phase Φ to be discontinuous at the boundary, where the coordinate is also discontinuous. It should be noted that these proposed nonlinear boundary conditions will not change the Hermitian character of the momentum operator. The change from linear conditions can be accommodated by phase rotations of the momentum operator in opposite directions at the boundary, which preserves Hermiticity. The character of the momentum operator as a matrix with nonlinear boundaries is developed in Appendix A.

Since each eigenfunction in Eq. 2 has a constant amplitude and constant phase gradient, the nonlinear boundary conditions are met for all real n and k, even though Φ itself is not gauge invariant. Thus the set of eigenfunctions selected by the boundary is the same continuous spectrum as for the infinite line discussed above. This means that all real values are allowed for the eigenvalue n, and thus gauge invariance is restored, just as for the infinite line. Of course, with integration over the finite domain of the ring coordinate, the set of eigenfunctions with a continuous spectrum loses its orthonormality.

In other words, it cannot be assumed that for the ring all eigenfunctions with real eigenvalues will form a Hilbert space. It is known, for example, that when two or more eigenfunctions with the form of Eq. 2, with different values of n are put in superposition, the resulting probability density will have periodic variation, and not all values of n will permit matching the period to that of the coordinate ring. To make this explicit, suppose one eigenfunction is chosen as the initial state of the ring, say with arbitrary real momentum eigenvalue q. Then we can systematically go through all the other eigenfunctions, add them all to that initial state, and ask which subset of them will result in periodicity of the probability density that matches the ring. The superposition will look like

$$\psi(x) = A e^{ikx} e^{iqx} \sum_{n=-\infty}^{n=+\infty} a_n e^{inx} \qquad \text{Eq. 4}$$

where k is still the gauge variable, and n is still the set of reals *a priori*. q is selected from the real continuous spectrum of momentum eigenvalues, and the set $a_n$ is composed of complex coefficients.

Write out the superposed eigenfunctions starting with lowest order term, n=0, which was the assumed initial state of the problem:

$$\psi(x) = Ae^{ikx}e^{iqx}(a_0 + a_1e^{in_1x} + a_2e^{in_2x} + \cdots) \qquad \text{Eq. 5}$$

where A is a real normalization constant, and the $a_j$ are complex coefficients. The $n_j$ are taken *a priori* to be an arbitrary sequence of real numbers in increasing order. We are looking for a subset of $n_j$'s that will allow $\psi^*\psi$ to be periodic on the interval from $-\pi \leq x \leq \pi$.

The complex conjugate of the above superposition is

$$\psi^*(x) = Ae^{-ikx}e^{-iqx}(a_0^* + a_1^*e^{-in_1x} + a_2^*e^{-in_2x} + \cdots) \qquad \text{Eq. 6}$$

Consider the product cross terms of Eq. 5 and Eq. 6 that contain terms with the zero-th order coefficients $a_0$ and $a_0^*$:

$$\psi^*\psi = A^2(a_0 a_1^* e^{-n_1x} + a_0^* a_1 e^{in_1x}) + (a_0 a_2^* e^{-n_2x} + a_0^* a_2 e^{in_2x}) + \ldots \qquad \text{Eq. 7}$$

Each of these zero order terms can be written as periodic real expressions such as:

$$(a_0 a_1^* e^{-n_1x} + a_0^* a_1 e^{in_1x}) = 2|a_0||a_1|\cos(n_1x + \Phi_{01}) \qquad \text{Eq. 8}$$

where $\Phi_{01}$ is the phase angle of the complex number ($a_0 a_1^*$). Each term will have the needed periodicity only if the $n_j$ are integers. Moreover, since the higher order cross terms will be functions of differences between these integers, those terms will also be properly periodic. Therefore, the general wave-function with superposed eigenfunctions can be written as Eq. 4, with n restricted to the set of integers, while k and q remain arbitrary reals. Also, the individual terms of eq. 4, with n restricted to integers form an orthonormal set and a conventional quantum Hilbert space. In fact, since the measured kinetic energy of this system is proportional to the square of $(q + n)$, with q continuous and n discrete, the Hilbert space will be that of a free particle in a system of quadratic energy bands.

## Comparison to the Weinberg nonlinearity.

The well known Weinberg nonlinearity[2] is thought to conflict with causality. Essentially, this is because that class of nonlinearity mixes eigenvectors and changes their directions in Hilbert space. The nonlinear boundary conditions discussed here will not have this property. Cross terms will be generated at the boundary from the $\psi^*\psi$ nonlinearity, but these only contribute to eigenfunction selection, and do not feed back into the eigenfunctions themselves. The dynamic variable remains the complex wave function evolving according to the still perfectly linear Schrödinger equation.

## Summary

Because gauge invariance apparently requires nonlinear boundaries in the one dimensional ring, there is a continuous infinity of non orthonormal momentum eigenfunctions with a continuous spectrum of real gauge independent eigenvalues. A discrete subset of these eigenfunctions is allowed in orthonormal superposition by the nonlinear boundary conditions. This subset is what conventional quantum mechanics would select as the complete set, but without gauge invariance and only a discreet spectrum. The subset remains superposable such that a ray in Hilbert space, once set up, is unperturbed by the boundary nonlinearity.

The impact of this work is least on the physics of the one dimensional periodic system for which experimental differences may be difficult to distinguish. However, this very straightforward mathematical presentation opens quantum mechanics to the nonlinear reality that permeates Newtonian mechanics and general relativity.

The necessity of nonlinear boundaries for this straightforward system suggests that nonlinearity may be important for the description of a quantum system coupled to its environment, possibly along the line of Fermi[3] Kostin[4] N. Gisin[5], Davidson[6], and others. Certainly the Josephson junctions used in some important quantum computer experiments[7] need to be re-examined, since these devices can be modeled as a particle on a ring such as considered here. It may be noted that Josephson junctions in the quantum limit have been assumed to have discrete energy levels.[8] This work predicts that these junctions will be found to have energy bands instead.

# Appendix A

## Nonlinear Hermitian Boundary Condition for the Momentum Operator

It has been shown in this article that linear boundary conditions for the momentum operator on a 1D ring produce gauge dependent eigenvalues, which must be unphysical. Therefore either the ring is not a valid quantum domain or the linear boundary conditions must be wrong. The appendix lays out some details of the alternative nonlinear and inhomogeneous boundaries.

The article showed that gauge invariance was restored by continuity at the boundary of the wave function amplitude and of the gradient of the phase. These nonlinear boundary conditions added a continuous spectrum of eigenvalues to a modified but still orthonormal and superposable Hilbert space. The appendix will show that the momentum operator with nonlinear boundaries still works in the usual operator eigenvalue equation for momentum.

The plan here is to start by presenting the matrix form of the momentum operator in the spatial representation for linear periodic boundary conditions. By "spatial representation" is meant that the system vector is represented by values of the complex wave-function at different spatial locations around the ring. In addition to linearity, standard quantum mechanics requires Hermiticity, which means the matrix must be square, with real diagonal elements, and other elements such that transposition of rows and columns are complex conjugates. The various matrix elements can then be chosen to correspond in the limit of small coordinate differences to the usual differential momentum operator.

Next, it will be considered how this particular matrix momentum operator would change because of the proposed nonlinear boundary conditions. It will be argued that the only change necessary is a rotation of the operator in two particular cells such that Hermiticity is preserved. The degree of rotation in those two cells of the operator depends on the phase of the wave-function. This dependence of the two cells on the wave function makes the operator nonlinear in principle, but preserves the essential structure of Hilbert space.

Finally, a commercial matrix solver[a] was used to find the eigenvalues as the nonlinear boundary condition was varied in a 20 by 20 momentum operator matrix. The predicted continuous and discrete parts of the eigenvalue spectrum emerged, directly confirming consistency in the use of nonlinear boundaries.

## The linear periodic momentum operator

The matrix form of the linear momentum operator in the spatial representation with linear periodic boundary conditions is shown below in Eq. a1. The gauge term is set to zero.

This operator is Hermitian. Its size, 7 by 7, reflects a division of the ring into 7 segments to be used as differential elements of the operator. The increment represented by one segment is *dx*. Next nearest neighbors are used to calculate the derivative so that the diagonal meets the requirement to be real, and the matrix as a whole to be Hermitian. Notice the *i* and *–i* in the (1,7) and (7,1) corners, which gives the operator linear periodic boundary conditions. The operator is both linear and homogeneous through the whole domain, including the boundary. A 7x7 matrix was chosen here simply for convenience of display. Later, calculations will be shown from a similar 20x20 matrix with nonlinear boundaries.

$$P_{op} = \frac{1}{2\,dx} \times \begin{vmatrix} 0 & -i & 0 & 0 & 0 & 0 & i \\ i & 0 & -i & 0 & 0 & 0 & 0 \\ 0 & i & 0 & -i & 0 & 0 & 0 \\ 0 & 0 & i & 0 & -i & 0 & 0 \\ 0 & 0 & 0 & i & 0 & -i & 0 \\ 0 & 0 & 0 & 0 & i & 0 & -i \\ -i & 0 & 0 & 0 & 0 & i & 0 \end{vmatrix} \qquad \text{Eq. a1}$$

## Change to nonlinear boundary conditions

How does this matrix operator change for the proposed nonlinear boundary conditions? First, express the derivative of an arbitrary complex function ψ in terms of its amplitude A and phase α:

$$\frac{d\Psi}{dx} = (A_x + iA\alpha_x)e^{i\alpha} \qquad \text{Eq. a2}$$

where the x subscript denotes differentiation by the coordinate variable. Therefore, if the nonlinear boundary condition is that $A$, $A_x$, and $\alpha_x$ are smooth, then what is required is a way to differentiate across the boundary where there may be a discontinuity in α. The linear matrix operator $P_{op}$ in Eq. a1 differentiates across the boundary twice, once in the first row, and once in the last row, so the salient thing to do at the boundary of the nonlinear operator is to rotate the (1,7) and (7,1) cells in opposite directions by $\Phi = \Delta\alpha$, as shown below in Eq. a3:

$$P_{op}^{nl} = \frac{1}{2\,dx} \times \begin{vmatrix} 0 & -i & 0 & 0 & 0 & 0 & ie^{i\Phi} \\ i & 0 & -i & 0 & 0 & 0 & 0 \\ 0 & i & 0 & -i & 0 & 0 & 0 \\ 0 & 0 & i & 0 & -i & 0 & 0 \\ 0 & 0 & 0 & i & 0 & -i & 0 \\ 0 & 0 & 0 & 0 & i & 0 & -i \\ -ie^{-i\Phi} & 0 & 0 & 0 & 0 & i & 0 \end{vmatrix} \qquad \text{Eq. a3}$$

Here $\Phi = \Delta\alpha$ is the phase discontinuity of α at the boundary. So if the eigenfunction being operated on has phase qx, where q is real and x goes from –π to π, then $\Phi = 2\pi q$.

This makes the matrix operator for $P_{op}^{nl}$ in Eq. a3 inhomogeneous in the wave function, and since those two cells now depend on the phase of the wave-function, nonlinear as well. The derivative of the wave-function calculated this way will nonetheless be smooth and continuous everywhere except for the discontinuity in α, which is exactly what is needed to preserve gauge invariance.

## Eigenvalue solutions

A matrix of the form of Eq. a3, but 20 by 20 instead of 7 by 7, was diagonalized by the "eig" function in Matlab[a], for various values of Φ between zero and π. Figure a1 below plots the numerical value of a range of the "eig" solutions for the eigenvalues vertically against Φ on the horizontal axis.

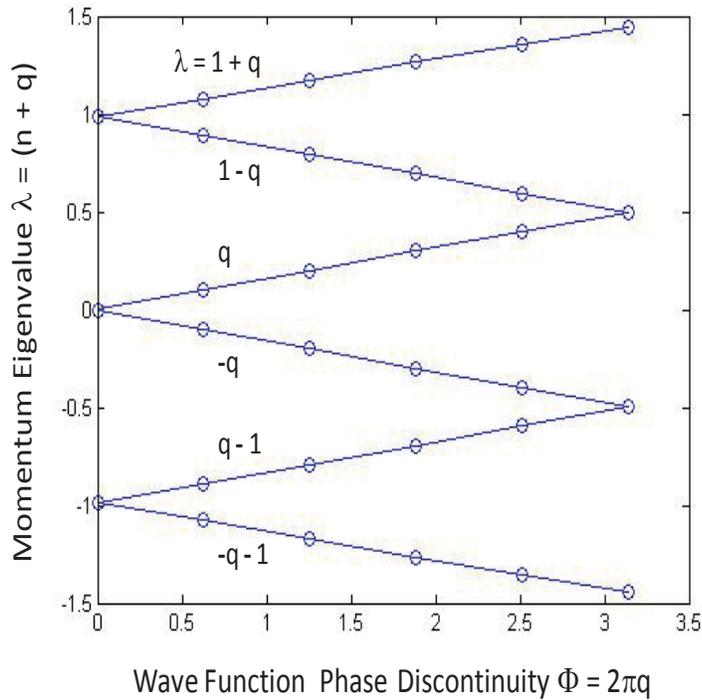

Wave Function Phase Discontinuity Φ = 2πq

If λ is the real eigenvalue of the operator, the solution for the nonlinear boundaries is λ= (q+n), where q is an element of the continuous and real eigenvalue spectrum, and n is an integer. This is precisely what emerged from the Matlab diagonalization. In contrast, linear boundary conditions would allow Φ = 0 only, and give only the quantized momentum values on the left vertical axis.

Figure a1. The lines represent the variation of the eigenvalues of the momentum operator above, as calculated by the Matlab "eig" function.[a] The magnitude of the phase discontinuity ϕ of the wave function is plotted on the horizontal axis, from 0 to π. The vertical axis is the momentum eigenvalue normalized to ℏ. When ϕ =0 on the left axis, the eigenvalues are discreet with the values -1, 0 and +1 shown. As ϕ increases, each eigenvalue splits into two momentum bands. An arbitrary state will have a single real value of q, but multiple n states in superposition.

## Appendix Endnote

[a] Matrix computations done with commercial package Matlab 2008a, The MathWorks, Natick, MA (2008).